\documentclass{aa}
\usepackage[varg]{txfonts}
\usepackage{graphicx}
\usepackage{txfonts}
\usepackage{natbib}
\usepackage{verbatim}
\usepackage[pdftex]{color}
\usepackage[dvipsnames]{xcolor}
\usepackage{hyperref}
\usepackage{ulem}
\usepackage{subfigure}
\hypersetup{
    colorlinks = true,
    citecolor ={blue}
}
\usepackage{mathrsfs}
\bibpunct{(}{)}{;}{a}{}{,}

\newcommand{\G}[1]{$Gaia$}
\usepackage{booktabs}
\definecolor{ultramarine}{rgb}{0.07, 0.1, 0.6} 
\definecolor{myblue}{rgb}{0.07, 0.2, 0.6} 
\definecolor{dopal}{rgb}{.70, .25, .05}

\begin{document}

\title{Revisiting the Galactic age–metallicity relation from wide white dwarf–main-sequence binaries 
}

\author{Alberto Rebassa-Mansergas\inst{1,2}\thanks{E-mail: alberto.rebassa@upc.edu}, Iset Tejero-Gómez\inst{1}, Roberto Raddi\inst{1}}
\institute{Departament de F\'isica, 
           Universitat Polit\`ecnica de Catalunya, 
           c/Esteve Terrades 5, 
           08860 Castelldefels, 
           Spain
           \and
            Institut d'Estudis Espacials de Catalunya (IEEC), C/Esteve Terrades, 1, Edifici RDIT, 08860 Castelldefels, Spain
           }

\date{Received ; accepted }

\abstract{
The  age–metallicity relation  (AMR)  is  a fundamental  observational
constraint for understanding the chemical  evolution of the Galaxy. As
reliable cosmochronometers,  white dwarfs in binary  systems with main
sequence companions (WD+MS binaries) provide excellent laboratories to
study this relation, since both  components are expected to be coeval.
}
{
We construct  a sample of  widely separated WD+MS binaries  using data
from  the  third  data  release  of  the  \G\,  mission  in  order  to
investigate the AMR of the Galactic disk.
}
{
The sample is identified using photometric measurements and parallaxes
of both  components.  White  dwarf ages  are derived  by interpolating
their  \G\,   absolute  $G$   magnitudes  and  BP-RP   colours  within
state-of-the  art  white  dwarf evolutionary  sequences.   We  compile
publicly available [Fe/H] abundances  for the main sequence companions
from  the  literature and  combine  them  using different  statistical
approaches  to  obtain  representative  metallicity  values  for  each
system.
}
{
We  derive  the  AMR  from  several  sub-samples  of  WD+MS  that  use
independent measurements of [Fe/H]  abundances and consistently find a
large dispersion  in [Fe/H]  at all ages.  This behaviour  agrees with
previous determinations  of the AMR  based on both WD+MS  binaries and
samples of isolated stars.
}
{
Our results reinforce  the observational evidence that the  AMR in the
Galactic   disk  exhibits   substantial   intrinsic  scatter,   likely
reflecting the combined effects of  multiple mechanisms such as radial
migration, inhomogeneous  chemical enrichment,  and variations  in the
star formation history.
}

\titlerunning{WD+MS binaries from \G\, DR3}
\authorrunning{Rebassa-Mansergas et al.}

\maketitle

\section{Introduction}
\label{introduction}

The relation between  stellar ages and chemical  abundances provides a
powerful observational  constraint on  the formation and  evolution of
the Milky  Way. In particular,  the age–metallicity relation  (AMR) of
Galactic disk stars encodes  information about the chemical enrichment
history of the interstellar medium  and the dynamical and evolutionary
processes  that  have  shaped  the  Galactic  disk  \citep{Twarog1980,
  Minchev2013}. Early observational  studies of nearby F  and G dwarfs
revealed  that the  AMR in  the solar  neighbourhood exhibits  a large
dispersion in metallicity at a  given age, challenging the predictions
of simple models of Galactic chemical evolution \citep{Edvardsson1993,
  Rocha-Pinto2000, Nordstrom2004}. Subsequent  analyses confirmed that
the Galactic disk  does not follow a  simple monotonic age–metallicity
sequence,  suggesting   that  processes  such  as   radial  migration,
variations  in  star  formation efficiency,  and  spatially  dependent
enrichment histories  play an important  role in shaping  the observed
distribution    of    stellar    ages    and    chemical    abundances
\citep{Haywood2008, Schonrich+Binney2009, Frankel2018}.

In recent years,  the combination of large  spectroscopic surveys with
the precise astrometric measurements provided  by the \G\, mission has
dramatically  improved  our  ability to  explore  the  chemo-dynamical
structure  of the  Milky Way.  Surveys  such as  APOGEE (Apache  Point
Observatory Galactic Evolution Experiment; \citealt{Majewski2017}) and
LAMOST  (Large Sky  Area Multi-Object  Fiber Spectroscopic  Telescope;
\citealt{Cuietal12}) now  provide chemical abundances for  hundreds of
thousands of  stars across large  regions of the Galactic  disk, while
\G\, parallaxes  enable improved determinations of  stellar parameters
and  ages.  These  data sets   have  revealed  a  complex  age–chemical
structure in the Galactic disk, with  the thin disk exhibiting a broad
metallicity  distribution  at  nearly  all ages  and  the  thick  disk
following  a   tighter  age–abundance   sequence  \citep{Feuillet2019,
  Queiroz2020, Sahlholdt2022,  Anders2023}. One  of the  most detailed
observational views of the Galactic  AMR has been obtained using large
samples of  sub giant stars with  precise age estimates, which  map the
age–metallicity  structure   of  the  disk  and   its  variation  with
Galactocentric radius \citep{Xiang+Rix2022}.

Despite major  progress enabled  by large  photometric, spectroscopic,
and  astrometric surveys,  robust  age  determinations for  individual
field      stars     remain      challenging     \citep{soderblom10-1,
  Tayar+Joyce2025}. Model  systematics and parameter  degeneracies can
produce uncertainties of several gigayears, even when population-level
trends are well defined \citep{Jorgensen2005, Ting2019, Curtis2020}.

White dwarfs  offer a complementary  approach to the  determination of
stellar  ages. As  the final  evolutionary  stage of  the majority  of
stars, white  dwarfs cool  in a predictable  manner over  time, making
them    reliable    cosmic    chronometers    \citep{Garcia-Berro1988,
  Fontaine2001}. Their cooling ages can be determined from atmospheric
parameters  and,  when  combined  with  estimates  of  the  progenitor
lifetimes  inferred  from  the initial–final  mass  relation,  provide
robust   determinations  of   the  total   ages  of   stellar  systems
\citep{Fouesneau2019,  Moss2022, Heintz2022}.  The precision  of these
measurements has improved  considerably with the advent  of \G,, which
provides accurate parallaxes and photometry for large samples of white
dwarfs.

Binary  systems  composed  of  a   white  dwarf  and  a  main-sequence
companion,   WD+MS   binaries,    constitute   particularly   valuable
laboratories for studying  the AMR. In wide systems  that have evolved
without mass transfer,  both stars are expected to share  the same age
and initial chemical composition. The white dwarf therefore provides a
reliable age  estimate through cosmochronology, while  the metallicity
of   the    system   can   be   measured    from   the   main-sequence
companion. Previous work demonstrated the  potential of this method by
deriving  the AMR  of the  solar neighbourhood  using samples  of wide
WD+MS  binaries identified  in spectroscopic  surveys \citep{Zhao2011,
  rebassa-mansergasetal16-2, rebassa-mansergasetal21-2}. These studies
found that the AMR inferred from  such binaries is consistent with the
large metallicity dispersion observed in field star samples.

In this work we further exploit the diagnostic power of WD+MS binaries
by  compiling publicly  available  metallicity  determinations of  the
companions and thus  increase the number of systems  with reliable age
and chemical measurements. By combining  precise white dwarf ages with
metallicity  measurements obtained  from these  multiple spectroscopic
surveys, we revisit the AMR in the solar neighbourhood from wide WD+MS
binaries.

The  paper  is organised  as  follows.  In Section\,\ref{s-sample}  we
describe  the sample  of  WD+MS  binaries analysed  in  this work.  In
Section\,\ref{s-feh} we  present the  compilation of  metallicities of
the main-sequence  companions. In  Section\,\ref{s-age} we  derive the
ages   of  the   systems   from  the   white   dwarf  components.   In
Section\,\ref{s-age-met}  we investigate  the resulting  AMR. Finally,
Section\,\ref{s-concl} summarises our main results and conclusions.

\section{The WD+MS sample}
\label{s-sample}

The WD+MS binary sample considered in  this work is an updated version
of the one we provided in \citet{rebassa-mansergasetal21-2}, using the
third data release  (DR3) of \G\, \citep{GaiaDR3} to  search for these
pairs. We started with a list  of white dwarf candidates identified by
\citet{gentile-fusilloetal21}, selecting  those with a  probability of
being a white dwarf greater than 50 percent and a parallax uncertainty
better than 10 percent. We  then searched for main sequence companions
within 100\,000  astronomical units from these  white dwarfs, ensuring
that  the  companions  had  good quality  data  by  applying  standard
astrometric  and   photometric  cuts  \citep{Gaia2018}.    The  search
resulted in the identification of  7\,249 WD+MS systems with distances
up to 500 parsecs.

In this  work, we  further restrict  the sample  to 4\,291  WD+MS with
assigned  spectral  types  to  their white  dwarfs,  classified  using
artificial  intelligence  methods \citep{Garcia2023,  Garcia2025}.  In
particular, the  former works apply  a Random Forest algorithm  to the
\G\,  BP/RP  spectral  coefficients  to  classify  white  dwarfs  into
different   types  with   high   accuracy   ($\simeq$91  percent   for
hydrogen-rich DA type white dwarfs; $\simeq$75 percent for non-DA type
white dwarfs).  Knowledge of the  spectral type is required  since the
cooling,  hence age,  of  a  white dwarf  depends  on its  atmospheric
properties. The fact that these white dwarfs have an assigned spectral
type  from  their  \G\,  spectra   alleviates  the  risk  of  using  a
probability  of  being  a  white  dwarf as  low  as  50  percent  from
\citet{gentile-fusilloetal19} to select candidates.

\section{Main sequence [Fe/H] abundances}
\label{s-feh}

We searched  for available [Fe/H] abundances  matching the coordinates
or \G,  IDs of the  main-sequence companions  in our sample  of 4\,291
WD+MS systems across different large-scale studies, which maximise the
number of matches relative to smaller  samples. This comes at the cost
of methodological  heterogeneity, as the abundances  are derived using
approaches  ranging   from  direct  spectroscopy  to   data-driven  or
machine-learning  estimates that  may not  rely on  the full  observed
spectra. This  caveat is  accounted for  in our  derivation of  AMR in
Section\,\ref{s-age-met}. The large-scale studies considered are:

\begin{itemize}
    \item  \citet{Ye2025},  who  develop a  catalogue  of  atmospheric
      parameters for 68 million stars  derived from \G\, BP/RP spectra
      by  fitting  synthetic  spectra   based  on  stellar  atmosphere
      models.  BP/RP fluxes  are corrected  using relations  involving
      stellar  colour,  magnitude,  and interstellar  extinction.  The
      claimed   [Fe/H]   precision   oscillates   between   0.12   and
      0.19\,dex. The search resulted in 532 matches.
    \item \citet{Huang2025}, who estimate  metallicities for about 100
      million stars from synthetic \G\, BP–RP and BP–G colours derived
      from corrected \G\,  BP/RP spectra. The precision  of the [Fe/H]
      estimates in this case varies from 0.05 to 0.25\,dex. The search
      resulted in 2\,758 matches.
    \item   \citet{Das2025},  who   derive   stellar  parameters   and
      abundances for nearly one million objects from their \G\, Radial
      Velocity Spectrometer  (RVS) spectra. They use  {\it The Cannon}
      \citep{Ness2015} to  transfer stellar parameters  and abundances
      from    the   GALAH    (GALactic   Archaeology    with   HERMES;
      \citealt{Martell2017})  data release  4  catalogue  to the  \G\,
      spectra.  The  achieved  [Fe/H]  precision  is,  in  this  case,
      0.02-0.1\,dex. The search resulted in 104 matches.
    \item \citet{Fallows2024},  who provide abundances for  28 million
      stars  applying an  Uncertain  Neural Network  model trained  on
      APOGEE  data  to  \G\,  BP/RP  spectra  and  selected  broadband
      photometric  colours  from  \G,,  2MASS and  WISE.   The  [Fe/H]
      precision  is  $\simeq$0.07\,dex.  The search  resulted  in  725
      matches.
    \item \citet{Hattori2025}, who estimates stellar metallicities for
      around   48  million   dwarfs   and  giants   situated  in   low
      dust-extinction regions by  applying tree-based machine-learning
      techniques to \G\, BP/RP spectra,  with training based on APOGEE
      data release  17 observations.  The claimed [Fe/H]  precision is
      $\simeq$0.09\,dex. The search resulted in 1\,769 matches.
    \item  \citet{Huang2022},   who  derive  stellar   parameters  and
      metallicities  for around  24 million  stars based  on SkyMapper
      Southern  Survey Data  Release 2  \citep[SMSS DR2;][]{Onken2019}
      and  \G\,  early  data  release  3  photometric  colours,  using
      training   data sets  composed   of   stars  with   spectroscopic
      parameters   from   earlier   surveys  at   different   spectral
      resolutions.   The   precision   of   the   [Fe/H]   values   is
      0.05–0.15\,dex. The search resulted in 284 matches.
\end{itemize}

As a result of this exercise we ended up with a sample of 2\,603 WD+MS
with  at  least two  [Fe/H]  abundances  as  reported from  the  above
studies.    According  to   the   Random   Forest  classification   of
\citet{Garcia2025}, 2\,070  of the white dwarfs  are hydrogen-rich DAs
($\simeq80$ percent), 187 are helium-rich DBs ($\simeq7$ percent), 309
are DCs  (no spectral features  in their spectra;  $\simeq12$ percent)
and the  remaining 37 ($\simeq1$  percent) belong to other  exotic and
less common types.  Thus, we restricted our sample to  the analysis of
DA systems  and excluded the  non-DAs, which  are more prone  to being
misclassified  by  the  Random   Forest  algorithm,  as  described  by
\citet{Garcia2025}. However, it has to  be emphasised that, as we will
see  in  Section\,\ref{s-age-met},  not  all these  2\,070  WD+MS  are
suitable for acquiring a reliable AMR.

\section{White dwarf ages}
\label{s-age}

We   use  the   same  approach   as  \citet{rebassa-mansergasetal21-1,
  rebassa-mansergasetal23} to derive the ages  of the white dwarfs. We
refer  the  reader to  these  works  for  a  full description  of  the
methodology and provide here a brief summary.

The total  age of a  given system is  calculated by summing  the white
dwarf cooling  age with  the main  sequence progenitor  lifetime. Both
parameters are  derived by  interpolating the  white dwarf  \G\, DR\,3
absolute  $G$-band magnitudes  and the  BP-RP colours  in the  cooling
sequences developed by the La  Plata group for different metallicities
\citep{Althaus2015,  Renedo2010,  Camisassa2016,  Camisassa2019}.  The
evolutionary sequences  cover the  full life  cycle of  the progenitor
stars, from the zero-age main sequence  to the white dwarf stage. They
therefore self-consistently incorporate an initial-final mass relation
\citep{Miller2016}  in  agreement with  semi-empirical  determinations
(e.g.  \citealt{Catalan08, Cummings2018}).  The sequences  account for
all major energy sources affecting  white dwarf cooling, including the
release of  latent heat  during crystallization and  the gravitational
energy associated with  phase separation. It is worth  noting that the
photometric data were corrected  for interstellar extinction using the
updated  3D   map  of  \citet{Lallement2014}  before   being  used  to
interpolate in the  theoretical sequences. We also  adopted a standard
$R_V =  3.1$, monochromatic extinction law  \citep{fitzpatrick2019} to
scale the 3D extinction into the three {\it Gaia} bands.

\begin{figure*}
    \centering
    \includegraphics[width=\columnwidth]{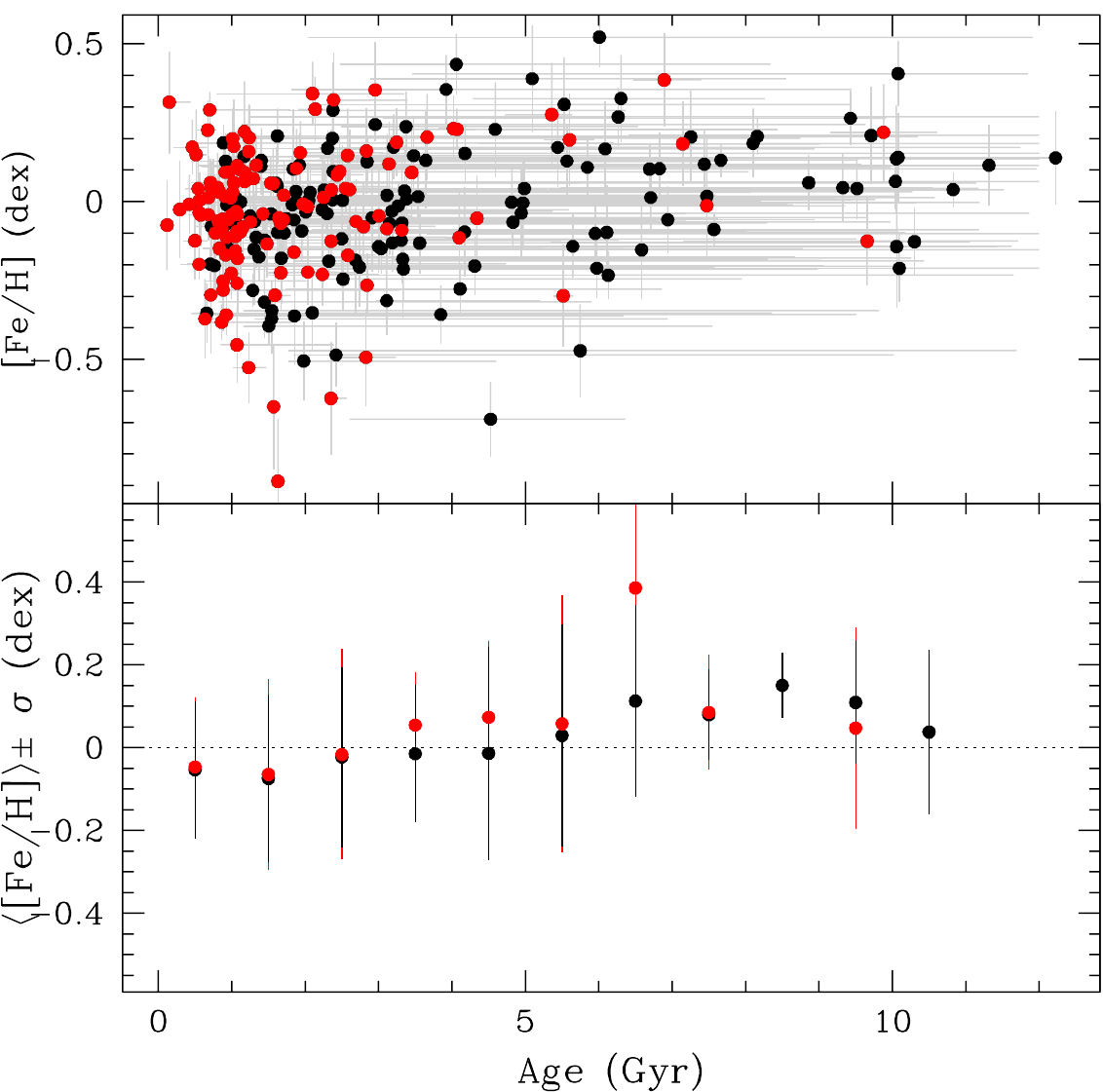}
    \includegraphics[width=\columnwidth]{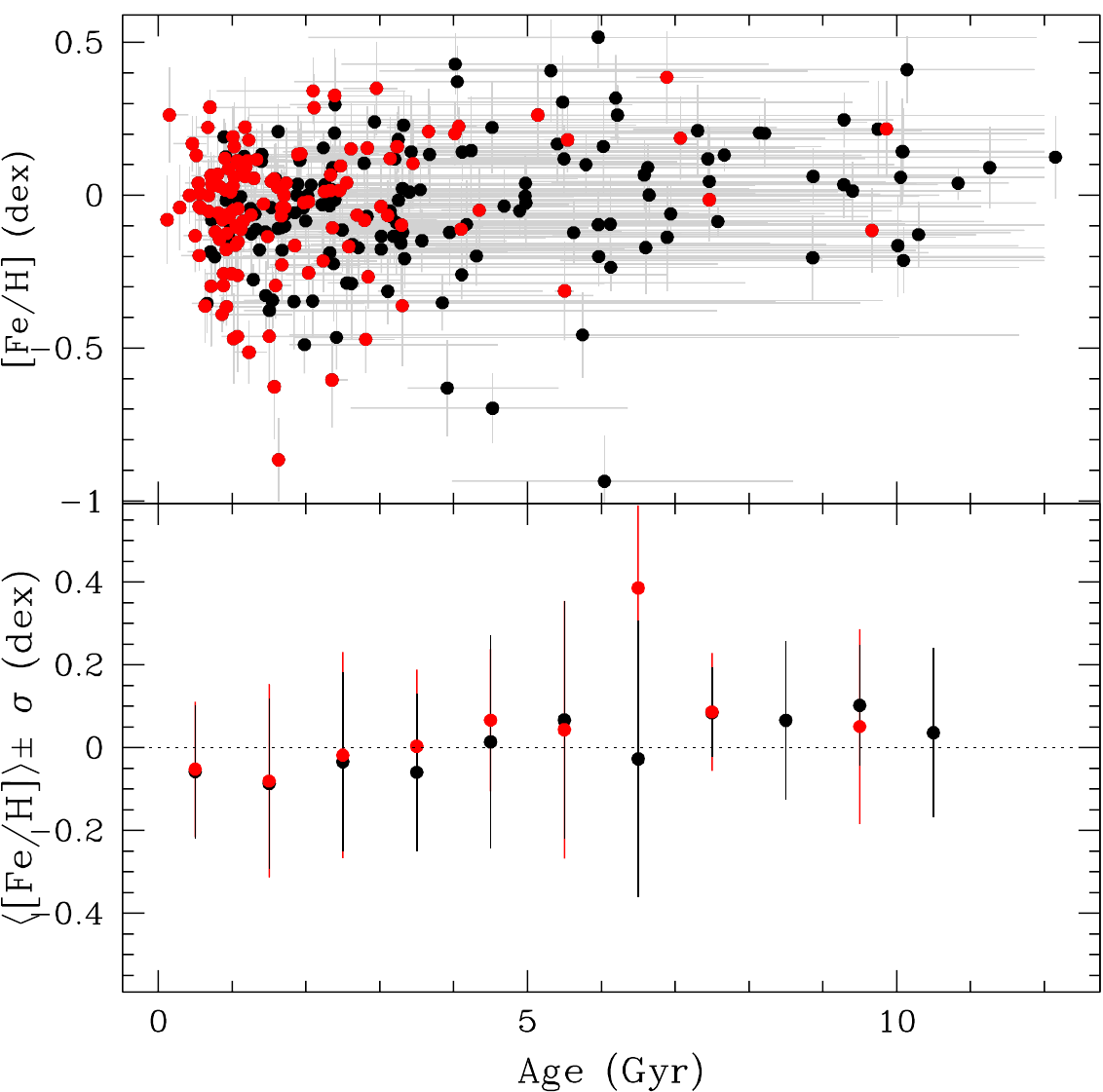}
    \caption{Left panels:  the AMR  (top) and the  average metallicity
      per age bin (bottom) arising from  the full sample of WD+MS with
      at  least  two available  [Fe/H]  measurements.  Red solid  dots
      indicate ages with relative errors  below 30 percent or absolute
      errors lower than 1.5\,Gyr. Right  panels: the same but applying
      a higher  weight to  [Fe/H] values  derived from  better quality
      data (see Section\,\ref{s-age-met-1} for details).}
    \label{f-m1}
\end{figure*}

\section{The age-metallicity relation}
\label{s-age-met}

In this section,  we determine the age-[Fe/H] relation  for our sample
using three different methods. Applying multiple approaches provides a
useful consistency check  and allows us to evaluate  how sensitive the
inferred relation  is to the  adopted methodology. Below,  we describe
each method and compare the resulting relations.

\subsection{The combined literature sample}
\label{s-age-met-1}

A first  estimate of  the AMR  is obtained  using the  full literature
compilation of  metallicity measurements  (Section\,\ref{s-feh}). That
is, for  the 2\,070 systems  containing DA white dwarfs  with multiple
[Fe/H]  determinations  available  in  the literature,  we  adopt  the
weighted mean  value, allowing  us to maximize  the number  of objects
while mitigating the impact of individual measurement uncertainties.

The weighted average of the measurements is:

\begin{equation}
    \langle [\mathrm{Fe}/\mathrm{H}] \rangle =
\frac{\sum_{i=1}^{n} w_i \,[\mathrm{Fe}/\mathrm{H}]_i}
{\sum_{i=1}^{n} w_i},
\end{equation}

\noindent    where    $n$    is   the    number    of    measurements,
$w_i=1/\sigma_i^{2}$ is the weight for  each measure, i.e. the inverse
variance, and  $\sigma_i$ is the  precision associated to  each value.
Although each  catalogue described in Section\,\ref{s-feh}  provides a
typical [Fe/H] precision, we  instead adopt a representative precision
derived from our  sample. Specifically, for each  catalogue we compute
the mean of the quoted uncertainties for the stars from that catalogue
that are included in our sample, and use this value as the precision.

We then  define the  uncertainty of  $\langle [\mathrm{Fe}/\mathrm{H}]
\rangle$ as  the quadratic  sum of the  internal and  external errors:
$\sigma_{\langle           [\mathrm{Fe}/\mathrm{H}]           \rangle}
=\sqrt{\sigma_\mathrm{int}^{2}+\sigma_\mathrm{ext}^{2}}$.

The  external  error represents  the  dispersion  among the  different
measurements,  reflecting  how  much  the measured  values  vary  with
respect to the weighted mean. It is defined as:

\begin{equation}
     \sigma_\mathrm{ext}=
\sqrt{\frac{\sum_{i=1}^{n} w_i\, (\,[\mathrm{Fe}/\mathrm{H}]_i-\langle [\mathrm{Fe}/\mathrm{H}] \rangle)^{2}}
{(n-1)\sum_{i=1}^{n} w_i}},
\end{equation}

The  internal  error  represents  the  theoretical  precision  of  the
weighted  mean.  It  is  calculated  from  the  uncertainties  of  the
individual measurements:

\begin{equation}
     \sigma_\mathrm{int}=
\sqrt{\frac{1}
{\sum_{i=1}^{n} w_i}},
\end{equation}

Given that in  most cases the number of available  measurements $n$ is
small (2-3), $\sigma_{\langle [\mathrm{Fe}/\mathrm{H}] \rangle} \simeq
\sigma_\mathrm{int}$.

From the  original sample  of 2\,070 WD+MS,  we only  considered those
with  $\sigma_{\langle   [\mathrm{Fe}/\mathrm{H}]  \rangle}<0.2$\,dex,
which reduced the number of  objects to 636. This threshold represents
a compromise  between excluding measurements with  large uncertainties
and maintaining  a statistically  meaningful sample size.  The drastic
reduction  of   objects  (2\,070   to  636)  reflects   the  different
methodologies for  measuring the  [Fe/H] abundances of  the considered
studies, which yield  incompatible values in many  cases. Moreover, it
is important  to note that for  350 white dwarfs their  derived masses
are  below 0.53\,M$_{\odot}$.   This is  the lowest  white dwarf  mass
limit   allowed   in   the   evolutionary   sequences   described   in
Section\,\ref{s-age}.  The  progenitors of lower-mass white  dwarfs do
not have time to evolve out of the main sequence within the age of the
Universe. Hence, these 350 are further excluded from our sample, since
we  do not  have a  determination of  their ages.  The origin  of such
low-mass  white  dwarfs  in  wide  binaries  will  be  analysed  in  a
forthcoming publication.

The resulting  AMR for our remaining  286 WD+MS is illustrated  in the
top-left panel  of Figure\,\ref{f-m1}. Black solid  dots represent the
entire sample; red solid dots a restricted selection for which the age
errors are either below 1.5\,Gyr or  have relative errors of less than
30 percent. We observe a  large dispersion of [Fe/H] abundances around
the solar value at all ages. Indeed, the average metallicities per age
bin  are  consistent  with $\langle  [\mathrm{Fe}/\mathrm{H}]  \rangle
\simeq0$\,dex,  as  it  can  be  seen  in  the  bottom-left  panel  of
Figure\,\ref{f-m1}.

It  is  worth noting  that  most  WD+MS,  especially those  with  more
reliable   ages,   are   concentrated   towards   younger   age   bins
($<$5\,Gyr).  This is  because our  sample is  magnitude-limited, thus
favouring the identification of  brighter, intrinsically hotter, white
dwarfs with shorter  cooling ages. Generally, only  those white dwarfs
with masses under $\simeq$0.6\,M$_{\odot}$  have associated ages above
5\,Gyr, due to their long main sequence progenitor lifetimes. However,
because  the  main-sequence  lifetime   is  highly  sensitive  to  the
progenitor mass,  even small  errors in the  masses of  low-mass white
dwarfs lead to significantly different main-sequence lifetimes via the
initial-to-final mass relation and, therefore, to substantially larger
uncertainties in the total ages.

Given that the  [Fe/H] abundances used in this work  are obtained from
studies that  make use  of different methodologies  and data  sets, we
explored an  alternative approach  to estimate  the AMR  via favouring
measurements obtained from better quality  data. Thus, we repeated the
above  exercise  incorporating a  confidence  parameter  $q_i$ to  the
corresponding weights as follows:

\begin{equation}
     w_i=
\frac{q_i}
{\sigma_i^{2}}
\end{equation}

We adopted $q=1$ for the [Fe/H] measurements of \citet{Das2025}, since
they  are   derived  from   RVS  \G\,   spectra  of   resolving  power
$\simeq$11,500.  For  the  [Fe/H]  determinations  of  \citet{Ye2025},
\citet{Hattori2025} and \citet{Fallows2024}, based  on the analysis of
low-resolution \G\, BP/RP spectra, we  used $q=0.7$. Finally, we fixed
$q=0.4$    for    the    abundances   from    \citet{Huang2022}    and
\cite{Huang2025},  which are  based  on photometric  colours. In  this
case, the internal error's definition was modified to:

\begin{equation}
     \sigma_\mathrm{int}=
\sqrt{\frac{\sum_{i=1}^{n} w_i^{2} \sigma_i^{2}}
{\sum_{i=1}^{n} w_i}}
\end{equation}

The  resulting AMR  and  the  average metallicities  per  age bin  are
illustrated in the top- and bottom-right panels of Figure\,\ref{f-m1},
respectively. It becomes  obvious that there are  no major differences
between the  results obtained from both  approaches. For completeness,
we tried varying  the associated $q$ values and  found similar results
in all cases.

A  comparison between  the AMRs  obtained  in this  Section and  those
provided              by             \citet{rebassa-mansergasetal16-2,
  rebassa-mansergasetal21-2}, which  also arise  from the  analysis of
WD+MS binaries, reveals very similar features. Namely, a large scatter
of [Fe/H] abundances at any given age.

\begin{figure*}
    \centering
    \includegraphics[width=\columnwidth]{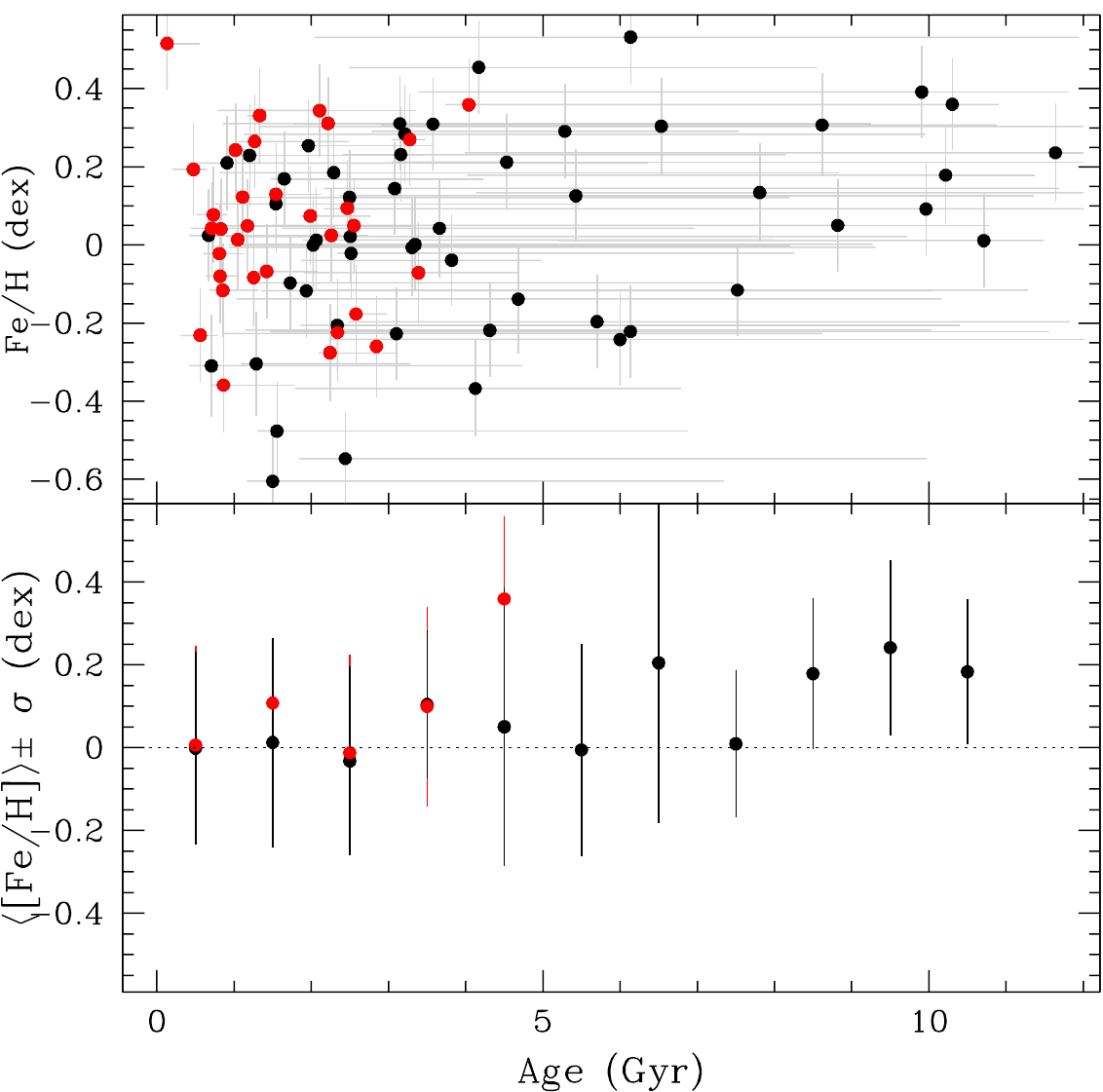}
    \includegraphics[width=\columnwidth]{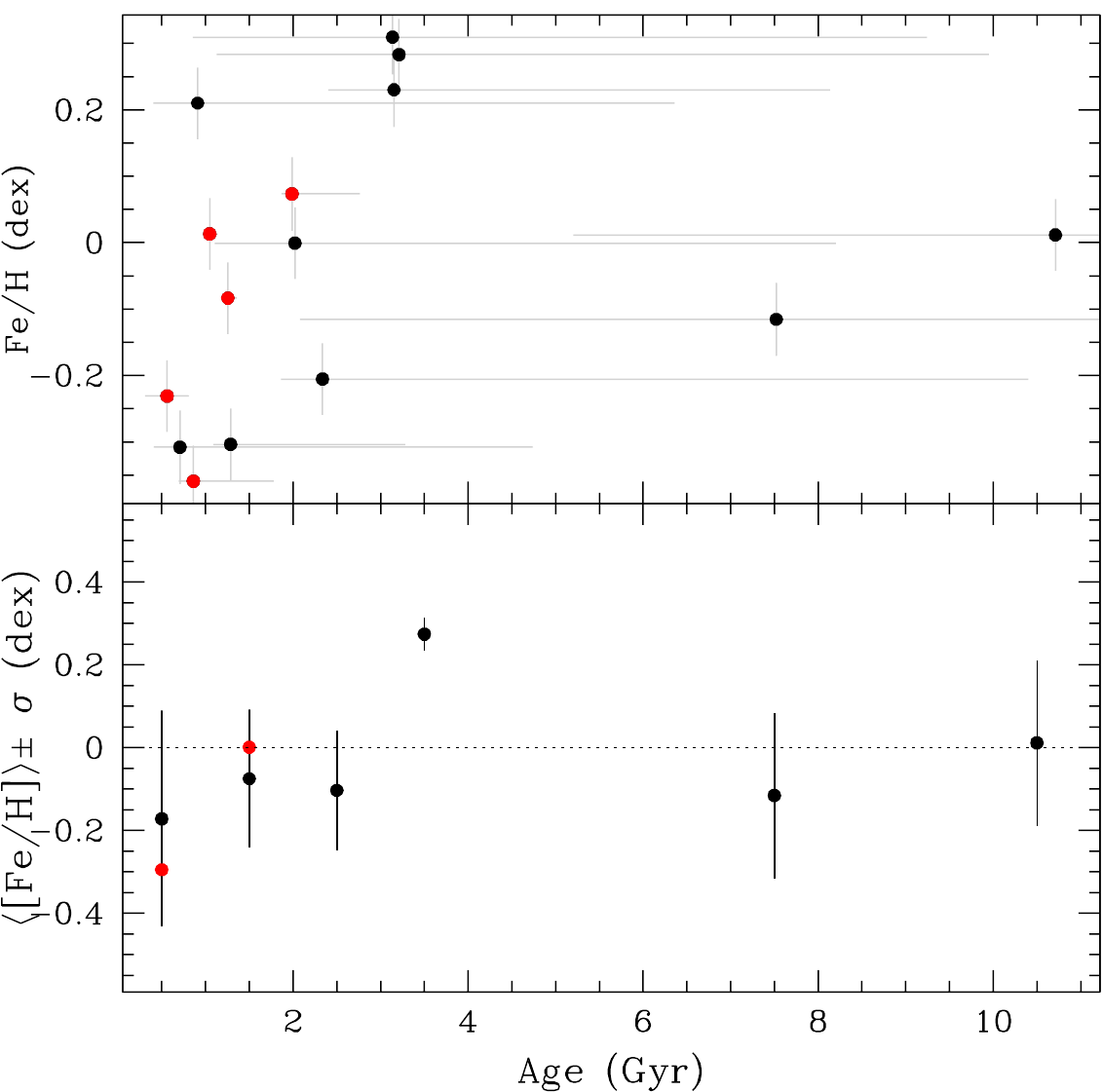}
    \caption{Left panels: the same as Figure\,\ref{f-m1} but for WD+MS
      with    [Fe/H]    abundances    from    \citet{Huang2022}    and
      \cite{Huang2025}    satisfying    our    $\tau<1.5$    threshold
      criterion.  Right panels:  the restricted  sample of  WD+MS with
      [Fe/H]  values  from   \citet{Huang2022},  \cite{Huang2025}  and
      \citet{Das2025}  also fulfilling  the $\tau<1.5$  criterion. See
      Section\,\ref{s-age-met-2} for details.}
    \label{f-m2-1}
\end{figure*}

\begin{figure*}
    \centering
    \includegraphics[width=\columnwidth]{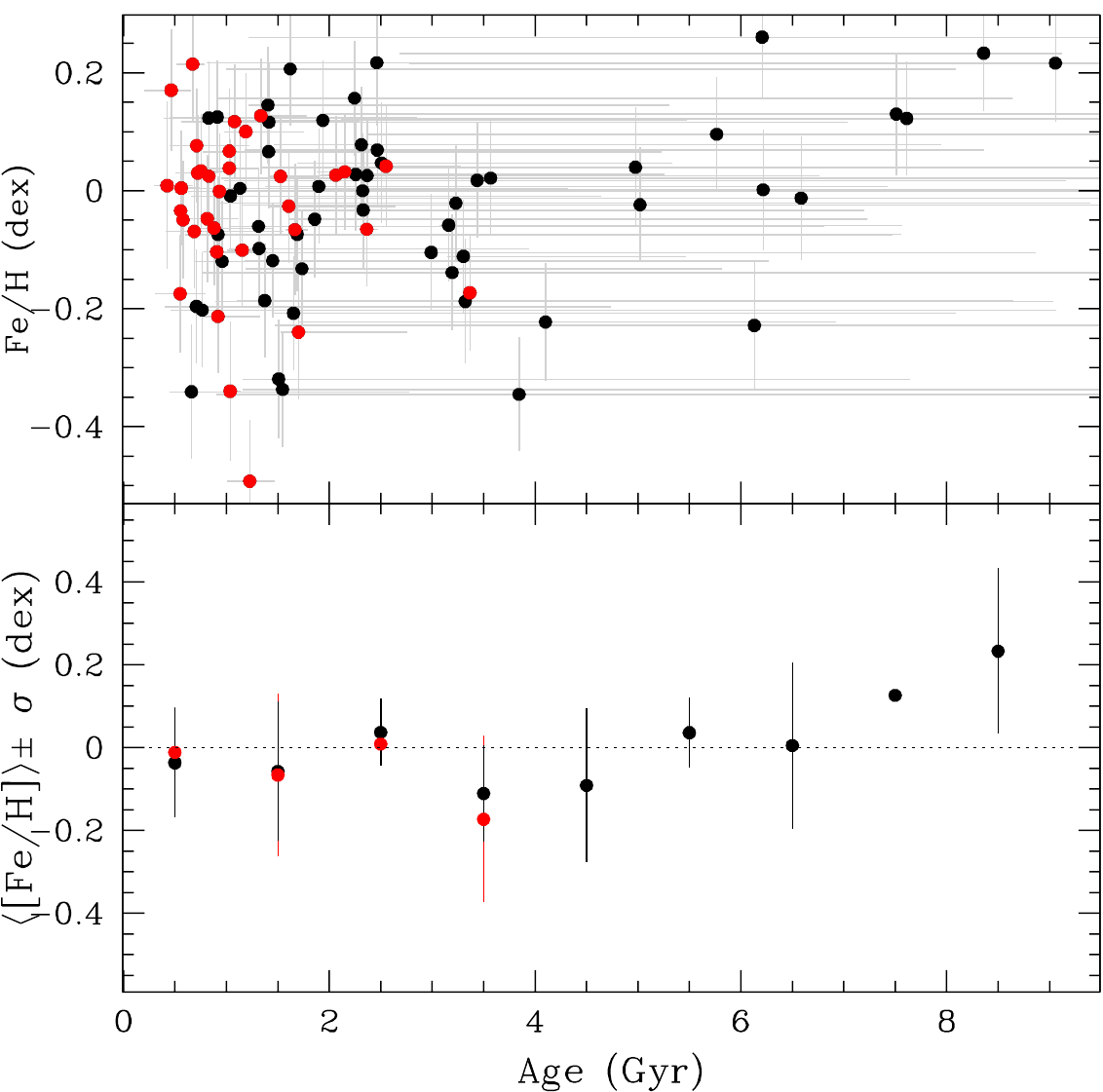}
    \includegraphics[width=\columnwidth]{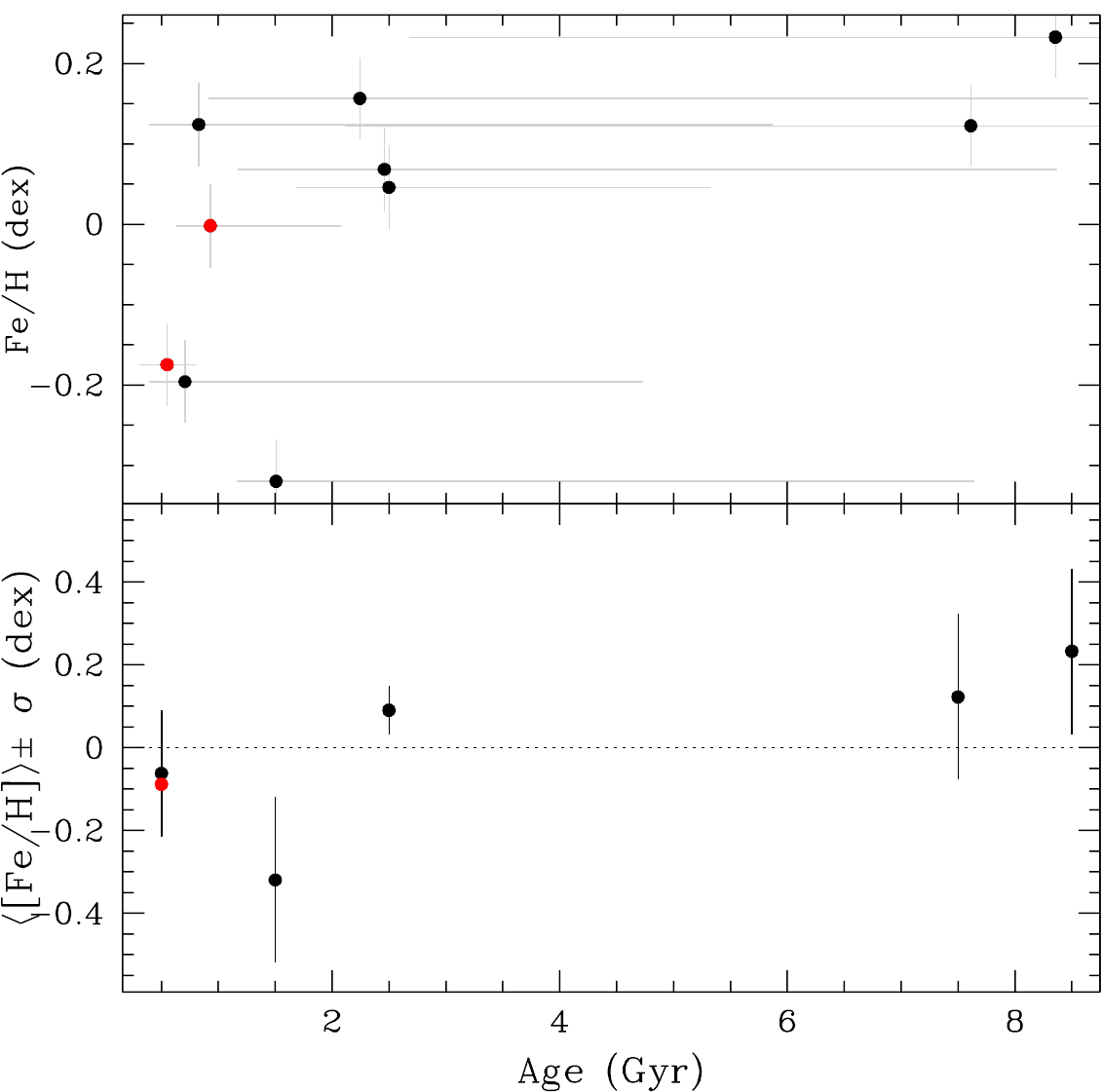}
    \caption{Left panels: the same as Figure\,\ref{f-m1} but for WD+MS
      with [Fe/H] abundances  from \citet{Ye2025}, \citet{Hattori2025}
      and  \citet{Fallows2024}  satisfying  our  $\tau<1.5$  threshold
      criterion.  Right panels:  the restricted  sample of  WD+MS with
      [Fe/H]   values    from   \citet{Ye2025},   \citet{Hattori2025},
      \citet{Fallows2024}  and  \citet{Das2025}  also  fulfilling  the
      $\tau<1.5$   criterion.   See   Section\,\ref{s-age-met-2}   for
      details.}
    \label{f-m2-2}
\end{figure*}

\subsection{The consistency-filtered sample}
\label{s-age-met-2}

To  assess  the impact  of  potential  systematic differences  between
literature measurements, we construct a more conservative sub sample by
retaining  only objects  whose  [Fe/H]  determinations from  different
studies agree within a specified  threshold. This procedure is applied
iteratively  when  additional  data sets  are  included,  progressively
reducing the  sample to  objects with mutually  consistent metallicity
measurements.

We begin  considering the  measurements derived from  \G\, photometric
colours   of   \citet{Huang2022}   and   \cite{Huang2025},   hereafter
[Fe/H]$_1$  and  [Fe/H]$_2$,  respectively.   We  obtain  the  average
metallicities  and  associated  uncertainties   in  the  same  was  as
described  in  Section\,\ref{s-age-met-1},   without  considering  the
confidence parameter $q_i$. However, we take into account only objects
satisfying the following threshold criterion:

\begin{equation}
     \tau=
\frac{|\,[\mathrm{Fe}/\mathrm{H}]_1-[\mathrm{Fe}/\mathrm{H}_2]\,|}
{\sqrt{\sigma_{\,[\mathrm{Fe}/\mathrm{H}]_1}^{2}+\sigma_{\,[\mathrm{Fe}/\mathrm{H}]_2}^{2}}}\,\,\,<\,\,1.5,
\label{eq-crit}
\end{equation}

\noindent     where    $\sigma_{\,[\mathrm{Fe}/\mathrm{H}]_1}$     and
$\sigma_{\,[\mathrm{Fe}/\mathrm{H}]_2}$ are the individual measurement
errors. The  selection criterion $\tau<1.5$ is  somewhat arbitrary and
represents  a   compromise  between   a  more   restrictive  threshold
(e.g.  $\tau<1$),  which  would  exclude  a  substantial  fraction  of
systems, and a more permissive one (e.g.  $\tau<3$), which would admit
systems with  less robust measurements. For  completeness, we explored
these alternative thresholds and found  that they do not significantly
affect the results,  except from the expected variation  in the number
of selected systems.  Thus, the $\tau <1.5$  restriction was fulfilled
by 84  of the 116 WD+MS  with common metallicities and  available ages
from both works  and the resulting AMR and average  [Fe/H] per age bin
are illustrated in the left panels of Figure\,\ref{f-m2-1}.

We further restricted the sample  by forcing the average [Fe/H] values
obtained  above   to  surpass  the   same  criterion  as   defined  in
Equation\,\ref{eq-crit}    but    with     the    measurements    from
\citet{Das2025}. We recall the reader that these [Fe/H] abundances are
obtained  from  \G\, RVS  spectra  and  are  hence the  most  reliable
measurements. Only 15  WD+MS were left after this process  and the AMR
arising   from    them   is   shown    in   the   right    panels   of
Figure\,\ref{f-m2-1}.

We continue  this analysis considering the  [Fe/H] abundances obtained
from   the   \G\,   BP/RP    spectra   provided   by   \citet{Ye2025},
\citet{Hattori2025}  and \citet{Fallows2024}.  Of the  115 WD+MS  with
common metallicities  among the three  studies and available  ages, 90
have   values   that   satisfy  the   $\tau<1.5$   criterion   between
\citet{Ye2025}  and  \citet{Hattori2025},  and 89  between  the  three
considered works. The resulting AMR can  be seen in the left panels of
Figure\,\ref{f-m2-2}.  We  further  apply the  criterion  of  Equation
\ref{eq-crit} to  the 89  mean [Fe/H] values,  this time  adopting the
measurements from \citet{Das2025}, thereby restricting the sample. The
AMR  resulting from  the just  10 WD+MS  fulfilling this  selection is
shown in the right panels of Figure\,\ref{f-m2-2}.

For  completeness,  we  also  evaluated how  many  WD+MS  fulfill  the
$\tau<1.5$    criterion    when    considering   all    six    studies
(\citealt{Ye2025,   Huang2022,    Huang2025,   Das2025,   Fallows2024,
  Hattori2025}).  Only  five  objects  satisfied  the  selection  and,
consequently, we do not show the AMR.

The  results  obtained  in  this  section reveal  that  using  a  more
conservative sub-sample yields results that  are very similar to those
obtained      from      the      full      literature      compilation
(Section\,\ref{s-age-met-1}).

\begin{figure}
    \centering
    \includegraphics[width=\columnwidth]{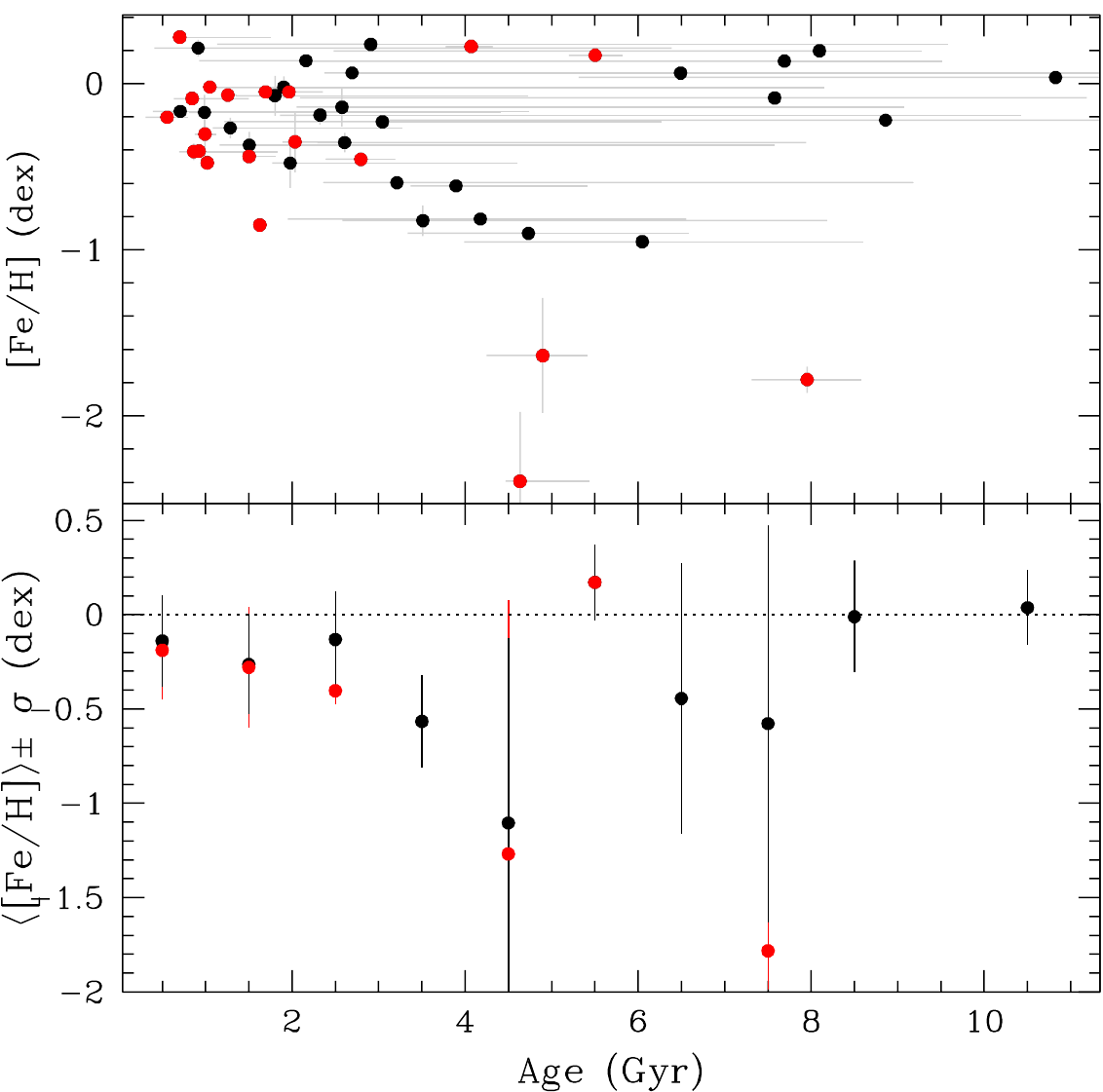}
    \caption{The same as Figure\,\ref{f-m1}  but for WD+MS with [Fe/H]
      abundances from our reference sample \citep{Das2025}.}
    \label{f-m3}
\end{figure}

For completeness, we show in  the Appendix the comparisons between the
[Fe/H] measurements of the considered studies throughout this section.

\subsection{The adopted reference sample}

Finally,  we   derive  the  AMR  using   [Fe/H]  measurements  adopted
exclusively  from   the  most   reliable  literature  source   in  our
compilation  \citep{Das2025}. This  approach  provides an  independent
estimate  of  the relation  that  avoids  combining measurements  from
heterogeneous analyses.

47  WD+MS in  our sample  have [Fe/H]  values from  \cite{Das2025} and
available ages. The AMR can be seen in Figure\,\ref{f-m3} and displays
the    typical   scatter    of    ages   found    in   the    previous
sections.  Interesting is  the  apparent  overabundance of  metal-poor
objects with ages between 3.5--7\,Gyr,  two of which are associated to
large   uncertainties   in   their  metallicities   and   are   likely
unreliable. This  feature is likely due  to the low number  of objects
considered.

\section{Summary and conclusions}
\label{s-concl}

In this work  we have presented the age-metallicity  relation, ARM, in
the solar neighbourhood as revealed from different sub-samples of wide
white dwarf  plus main sequence,  WD+MS, binaries. The  metallicity is
traced  by the  [Fe/H]  abundances of  the  main sequence  companions,
publicly  available  from  the  literature  \citep{Ye2025,  Huang2022,
  Huang2025, Das2025,  Hattori2025, Fallows2024}, whilst the  ages are
derived from the \G\, photometry and astrometry of the white dwarfs.

Independently of the sample used, we find a large scatter of [Fe/H] at
any age,  in agreement with  our previous work dedicated  to constrain
the  relation  between  age   and  metallicity  using  WD+MS  binaries
\citep{rebassa-mansergasetal16-2, rebassa-mansergasetal21-2}.  This is
also   in  line   with  studies   that  make   use  of   single  stars
\citep{Holmbeg2009,     Casagrande2011,    Haywood2013,     Patil2023,
  Casamiquela2024}.

It is important to remark  that, despite the important contribution of
WD+MS binaries to this topic, some factors limit the robustness of the
results obtained so far.

First,  as  noted  above,   lower-mass  white  dwarfs  originate  from
progenitors with longer main-sequence  lifetimes. Because the lifetime
inferred  from the  initial-to-final  mass relation  (IFMR) is  highly
sensitive to progenitor  mass, even small uncertainties  in the masses
of  low-mass  white dwarfs  translate  into  large variations  in  the
estimated  main-sequence lifetimes  and,  consequently,  in the  total
ages. As  a result, a  significant fraction of our  age determinations
carry substantial uncertainties, especially at old ages.

Second,  the  [Fe/H]  abundances   analysed  here  are  derived  using
heterogeneous methodologies and are available for a relatively limited
sample.  This limitation  will  be alleviated  by  the 4MOST  (4-metre
Multi-Object Spectroscopic Telescope) survey \citep{deJong2022}, which
is beginning operations in mid 2026 and will target, through its White
Dwarf Binary  Survey, $\simeq$2500  main-sequence companions  to white
dwarfs  in common  proper-motion pairs,  as well  as several  thousand
WD+MS systems  in close  orbits \citep{Toloza2023}.  The substantially
larger  sample  size, together  with  the  homogeneous nature  of  the
observations  and analysis,  will enable  a more  robust and  detailed
investigation of the AMR.

Finally, there is  still no consensus on  the observational properties
of   the   IFMR   \citep{Cummings2018,   Marigo2020,   Cunningham2024,
  Ironi2025}.  The  inferred  main-sequence lifetimes  of  progenitors
depend sensitively on  the adopted IFMR, which  can vary significantly
\citep{rebassa-mansergasetal16-2}.  Wide  binaries with  independently
determined ages  for the non-degenerate companions  can help constrain
the   IFMR   \citep{Catalan08,    Barrientos2021};   however,   robust
constraints require large samples of WD+MS binaries hosting relatively
massive white dwarfs ($\simeq$0.7,M$_\odot$).  Such objects have short
progenitor  lifetimes  \citep{Camisassa2016, Camisassa2019},  allowing
the  total system  age  to be  well approximated  by  the white  dwarf
cooling  age,  which  is  more reliable  and  largely  independent  of
metallicity. However, massive white dwarfs are both intrinsically rare
and faint, owing to their small  radii and rapid cooling. As a result,
they  represent  only a  small  fraction  of  white dwarfs  and  WD+MS
binaries in  volume-limited samples  \citep{McCleery2020, Jimenez2023,
  Kilic2025},   and   are   also   more   difficult   to   detect   in
magnitude-limited   surveys  \citep{Rebassa2015a,   Torres2023}.  Even
within  4MOST, only  a small  fraction ($\simeq$10  percent) of  WD+MS
binaries are expected  to host massive white dwarfs.  In this context,
the  Legacy Survey  of Space  and  Time (LSST)  at the  Vera C.  Rubin
Observatory \citep{Ivezic2019}  will play a  key role by  enabling the
identification of this elusive population \citep{Rebassa2025ESO}.

\begin{acknowledgements}

We thank the referee for the comments.  This work was supported by the
Spanish MINECO grant PID2023-148661NB-I00 and by the AGAUR/Generalitat
de Catalunya  grant SGR-386/2021.  RR acknowledges  support from Grant
RYC2021-030837-I,  funded by  MCIN/AEI/  10.13039/501100011033 and  by
“European Union  NextGeneration EU/PRTR”.  This work  presents results
from the European Space Agency (ESA)  space mission \G..  \G, data are
being processed  by the \G\,  Data Processing and  Analysis Consortium
(DPAC). Funding for the DPAC  is provided by national institutions, in
particular  the institutions  participating in  the \G\,  MultiLateral
Agreement     (MLA).      The      \G\,     mission     website     is
\url{https://www.cosmos.esa.int/gaia}.   The \G\,  archive website  is
\url{https://archives.esac.esa.int/gaia}.
 
\end{acknowledgements}

\newpage

\begin{appendix}

\section{[Fe/H] abundance comparison}

\begin{figure}[h]
    \centering
    \includegraphics[width=0.8\columnwidth]{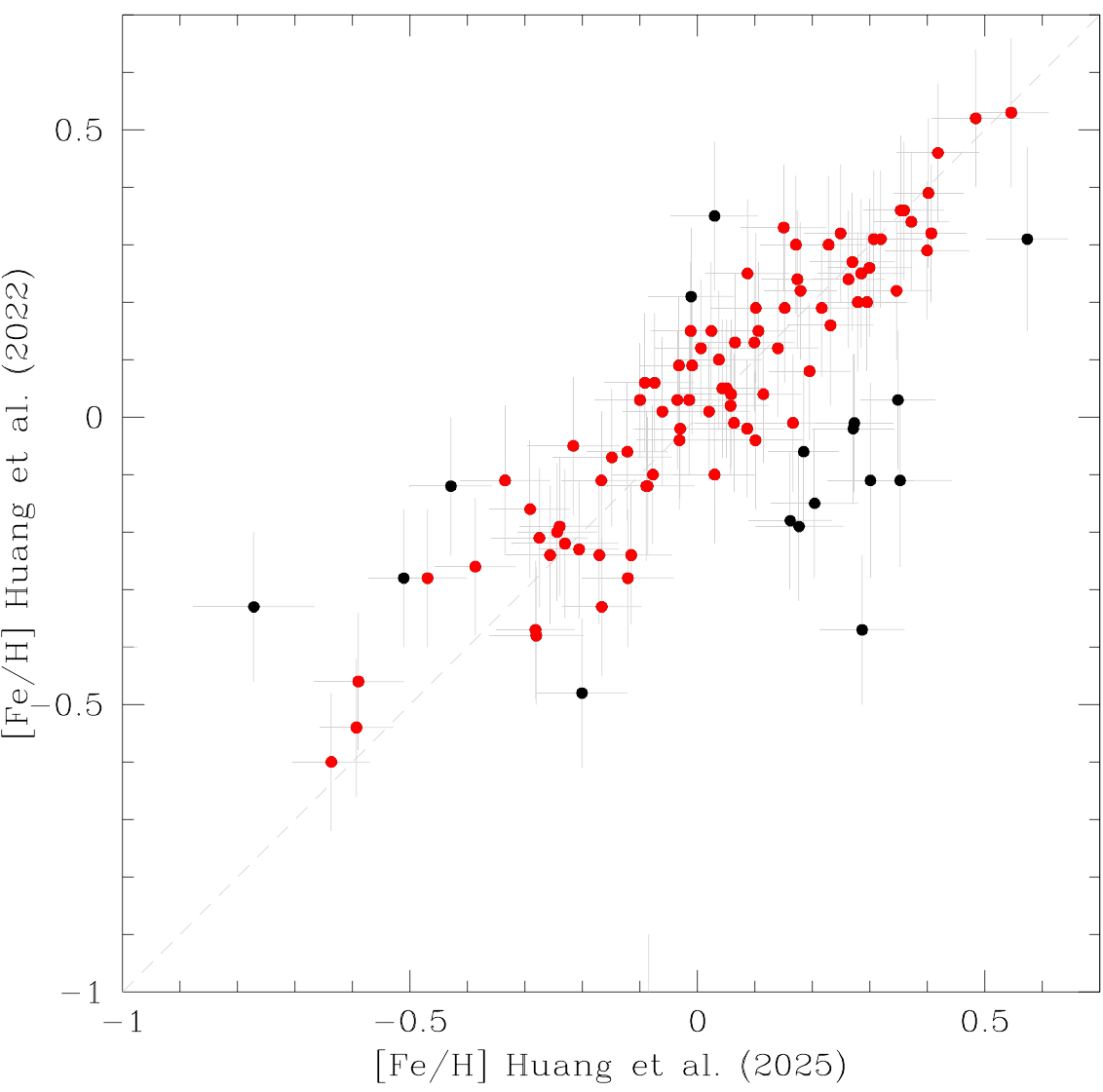}
    \includegraphics[width=0.8\columnwidth]{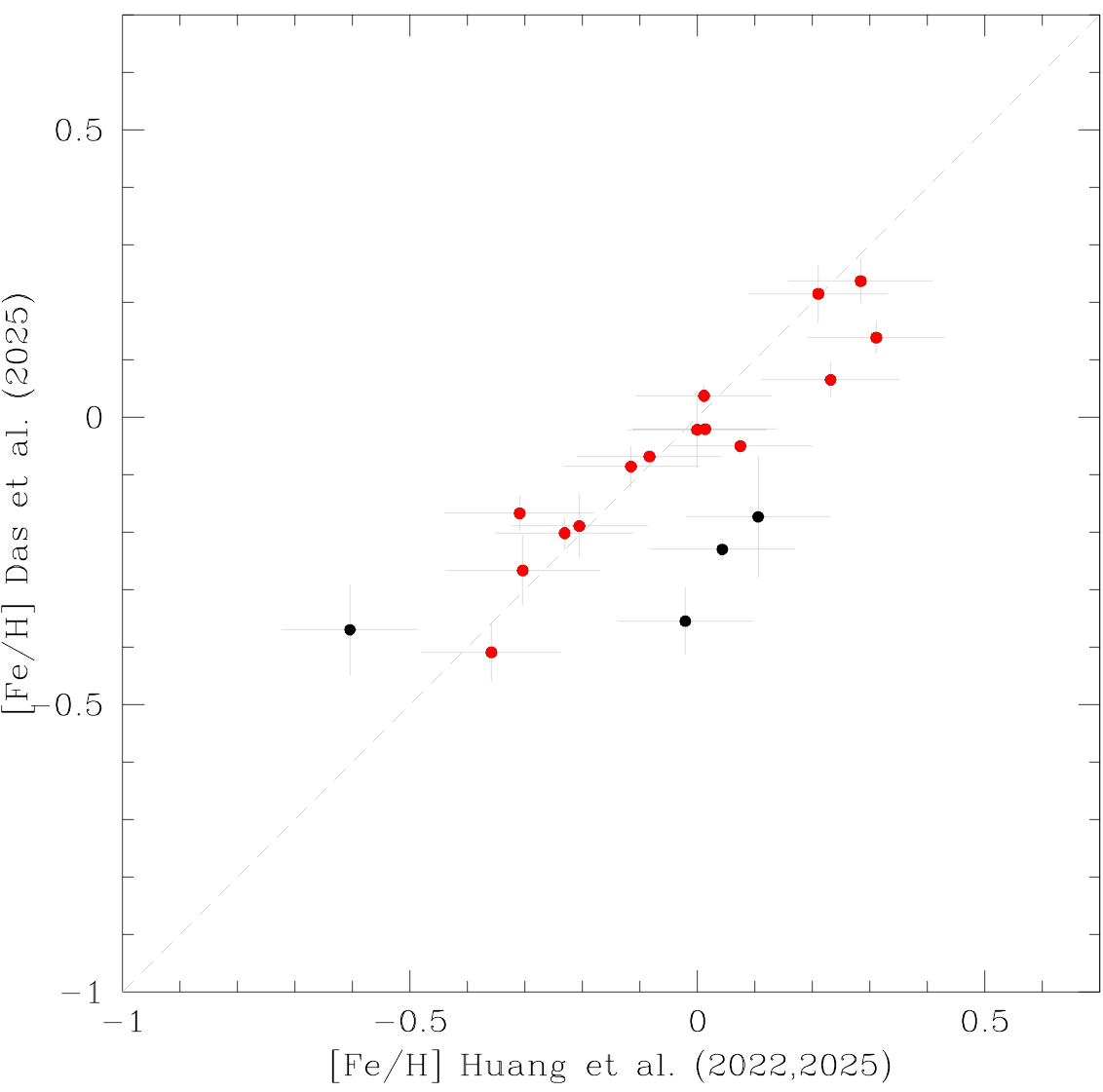}
    \caption{Top panel: comparison between  the [Fe/H] measurements of
      \citet{Huang2022} and \citet{Huang2025}.  Red solid dots fulfill
      the  $\tau<1.5$  condition.  Bottom  panel:  comparison  between
      average    [Fe/H]    values     from    \citet{Huang2022}    and
      \citet{Huang2025} fulfilling the $\tau<1.5$ condition versus the
      determinations  of  \citet{Das2025}.  Red solid  dots  represent
      again cases satisfying $\tau<1.5$.}
    \label{f-a1}
\end{figure}

\begin{figure}[h]
    \centering
    \includegraphics[width=0.8\columnwidth]{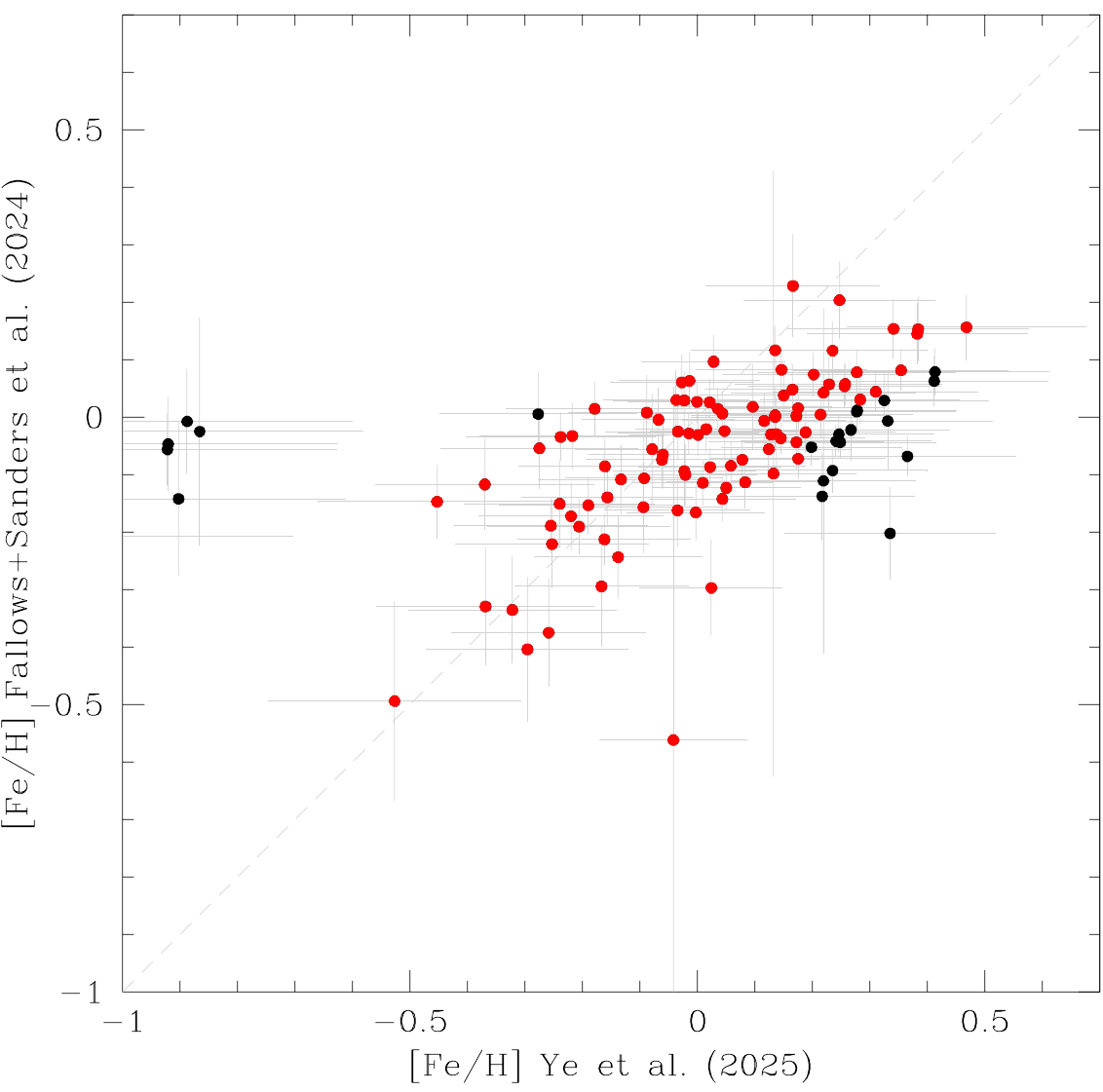}
    \includegraphics[width=0.8\columnwidth]{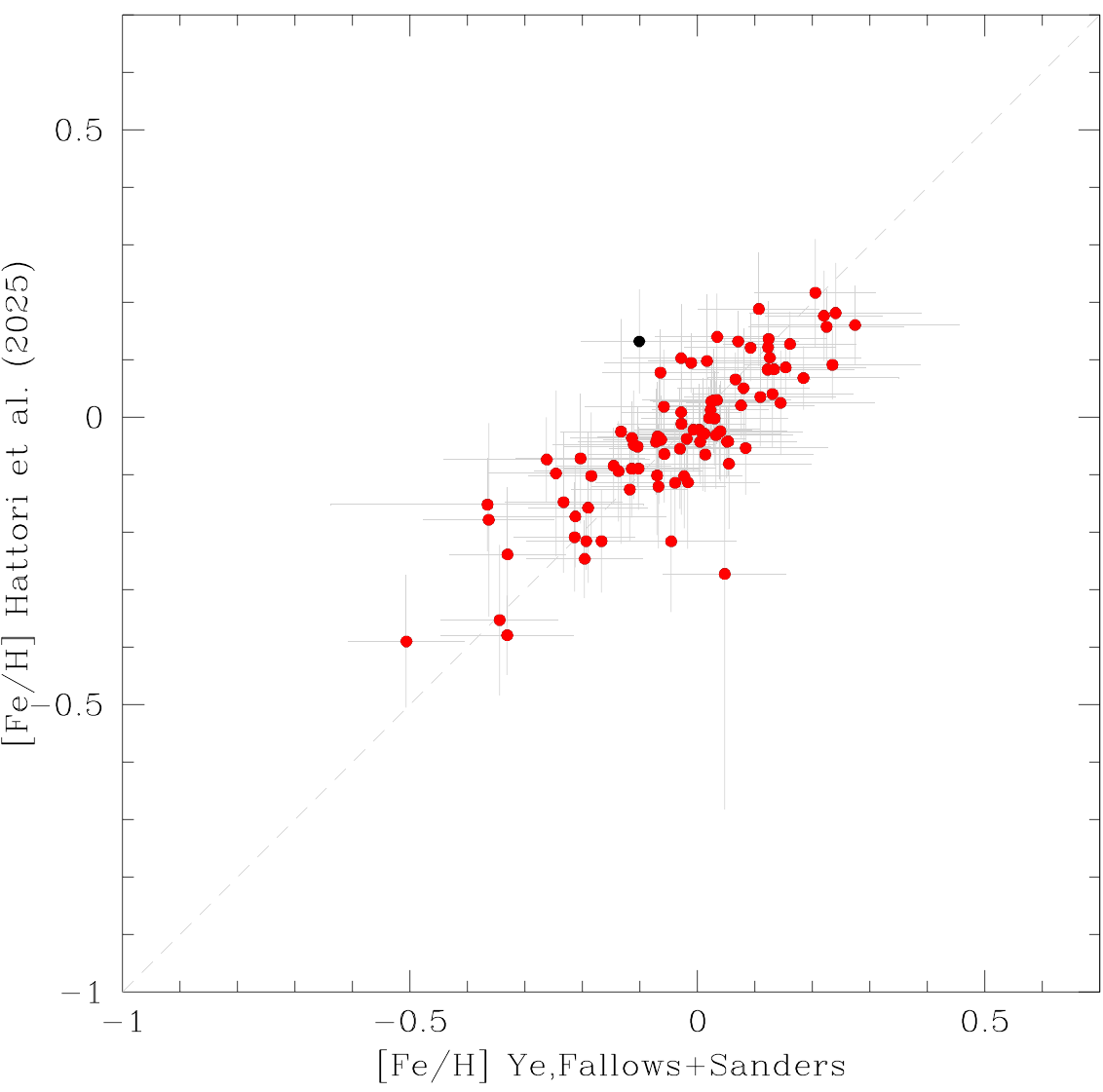}
    \includegraphics[width=0.8\columnwidth]{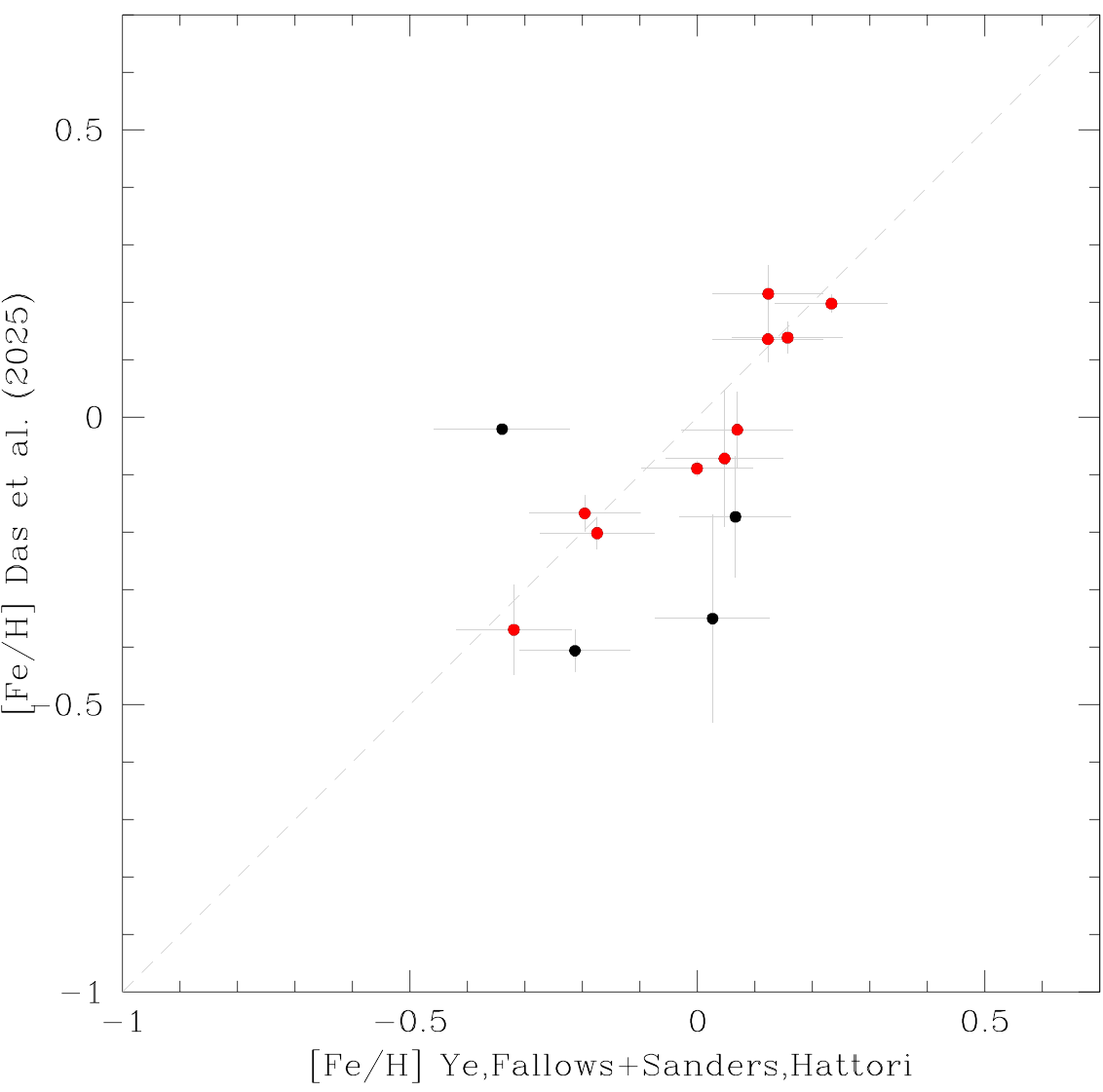}
    \caption{The same  as Figure\,\ref{f-a1} but comparing  the [Fe/H]
      values  from   \citet{Ye2025}  and   \citet{Fallows2024}  (top),
      \citet{Hattori2025} (middle) and \citet{Das2025} (bottom).}
    \label{f-a2}
\end{figure}

We provide in this Appendix  figures displaying the comparison between
the [Fe/H]  values measured from  the different studies  considered in
this            work,             as            detailed            in
Section\,\ref{s-age-met-2}.  Figure\,\ref{f-a1}  compares  the  values
obtained      by     \citet{Huang2022},      \citet{Huang2025}     and
\citet{Das2025}.  Figure\,\ref{f-a2} compares  the values  obtained by
\citet{Ye2025},    \citet{Fallows2024},     \citet{Hattori2025}    and
\citet{Das2025}.

\end{appendix}

\end{document}